\documentclass[aps,groupedaddress,showpacs]{revtex4-1}

\usepackage{amsmath,amssymb,bm,graphicx,epsfig,psfrag,ulem,color,natbib}
\usepackage{subfigure}
\usepackage{caption}
\usepackage{booktabs}

\newcommand{\bs}{\mathbf {s}}
\newcommand{\br}{\mathbf {r}}
\newcommand{\bv}{\mathbf {v}}
\newcommand{\bdr}{\mathbf {dr}}
\newcommand{\etal}{{\it et al.}~}

\begin{document}

\title{Large--scale superfluid vortex rings at nonzero temperatures}

\author{D.~H.~Wacks$^1$}
\email{daniel.wacks@ncl.ac.uk}
\author{A.~W.~Baggaley$^2$} 
\email{andrew.baggaley@glasgow.ac.uk}
\author{C.~F.~Barenghi$^3$} 
\email{c.f.barenghi@ncl.ac.uk}
\affiliation{$^1$School of Mechanical and Systems Engineering, Newcastle
University, Newcastle upon Tyne, NE1 7RU, UK,\\
$^2$ School of Mathematics and Statistics, University of Glasgow, G12 8QW,
Scotland,\\
$^3$ Joint Quantum Centre
Durham--Newcastle and School of Mathematics 
and Statistics, Newcastle University, Newcastle upon Tyne, NE1 7RU, UK}


\begin{abstract}
We numerically model experiments in which large--scale vortex
rings -- bundles of quantized vortex loops --
are created in superfluid helium by a piston--cylinder arrangement.
We show that the presence of a normal fluid vortex ring together
with the quantized vortices is essential to explain the coherence
of these large--scale vortex structures at nonzero temperatures,
as observed experimentally. Finally we argue that the interaction of
superfluid and normal fluid vortex bundles is relevant to recent
investigations of superfluid turbulence.
\end{abstract}


\maketitle

\section{Introduction}

Quantized vortex rings in superfluid liquid helium (helium~II)
have a special place \cite{BD-rings} in the vast vortex dynamics literature,
and have been recently visualized directly using tracer particles\cite{Bewley}.
The quantization of the circulation 
and the infinitesimal thickness of the vortex core make superfluid rings 
real--life examples of the (otherwise idealized) inviscid Euler 
dynamics which is described in textbooks. For example,
a well-known textbook result~\cite{Saffman}
is that the self--induced translational velocity $v_{self}$ of a thin-core 
vortex ring of large radius $R$, vortex core radius $a_0 \ll R$ 
and circulation $\Gamma$, moving in an incompressible inviscid fluid, 
is inversally proportional to $R$, and has the form:

\begin{equation}
v_{self}=\frac{\Gamma}{4 \pi R} \left [ \ln{\left (\frac{8R}{a_0} \right )} -\frac{1}{2} \right].
\label{eq:self}
\end{equation}

Eq.~(\ref{eq:self}) has been verified \cite{Reif} in superfluid helium.  
It must be noticed that in the helium context
$\Gamma$ and $a_0$ are not arbitrary parameters, but are fixed by
quantum mechanical constraints \cite{Donnelly}:
$\Gamma=\kappa=h/m \approx 10^{-3}~\rm cm^2/s$ (called the quantum of
circulation, where $h$ is Planck's constant and $m$ the mass of
one $^4$He atom), and $a_0 \approx 10^{-8}~\rm cm$ (proportional to
the superfluid coherence length). 

In superfluid liquid helium, vortex rings are usually created 
by high voltage tips.
The tips discharge ions which are accelerated by imposed electric
fields, until, upon reaching a critical velocity, 
quantized vorticity is nucleated \cite{Pomeau,Berloff,Adams}. 
The radius $R$ of the resulting rings is rather small, 
typically of the order of one micrometre only \cite{Golov}.

The more traditional piston--cylinder arrangement which is 
used to create much larger vortex rings in water or air has been implemented
in liquid helium only few times, but with remarkable results.
Borner \etal \cite{Borner1981,Borner1983,Borner1985} 
pushed helium through the circular hole
a cylindrical tube (diameter $D=0.8~\rm cm$) by the single stroke of a piston.
Using acoustic methods, they
determined size, position, velocity and circulation of the localised superfluid vortex structure which was ejected from the tube's hole.
They found that the structure travels a distance
which is at least 10 times its diameter $2R$
with reproducibly constant velocity and
circulation. They observed that the circulation
($2.3 < \Gamma <  4.85~\rm cm^2/s$ depending on the piston's action)
is much larger than $\kappa$. One infers that 
a large--scale macroscopic vortex ring was created, consisting of a large
number ($N = \Gamma/\kappa \approx  10^3$)
of individual, concentric, coaxial superfluid vortex rings  
travelling in the same direction -- essentially a compact bundle of
rings.
This interpretation was strengthened by further experiments
\cite{Stamm1,Stamm2}, 
particularly by Murakami and collaborators \cite{Murakami} 
who directly visualized the bundle using frozen hydrogen--deuterium
particles.

In a recent paper \cite{Wacks-PoF}, we discovered
that bundles of rings
moving in a perfect Euler fluid 
travel coherently over relatively large distances compared to their size.
During the evolution, the individual rings within a bundle move around
each other in a leapfrogging fashion.
Reconnections between individual  rings do not seem to
affect the speed and the coherence of the bundle (in other words, 
laminar and turbulent bundles move at approximately the same speed).
However, this work did not account for the
temperature independence observed by Borner \etal over the explored
temperature range $1.3 < T < 2.15~{\rm K}$. At nonzero temperatures
helium~II consists of two independent interpenetrating fluids: the inviscid
superfluid (with density $\rho_s$) and the viscous normal fluid (with
density $\rho_n$, where $\rho=\rho_s+\rho_n$ is the total density), which
affects the superfluid vortices via  friction force.
At $T=1.3~\rm K$ the normal fluid is certainly negligible \cite{DB}
($\rho_n/\rho=4.5 \%$), but at $T=2.15~\rm K$ it is certainly
not ($\rho_n/\rho=87.2 \%$). Therefore the friction \cite{BDV} 
should rapidly destroy a vortex ring over a short distance. In fact, the
range $\Delta z$ of a vortex ring of initial radius $R$ is \cite{BDV}
\begin{equation}
\Delta z \approx \frac{\rho_s \kappa}{\gamma}
\frac{\left(1-\gamma \gamma_0/(\rho_s^2 \kappa^2) \right)}
{\left( 1-\gamma_0'/(\rho_s \kappa) \right)} R,
\label{eq:range}
\end{equation}

\noindent
where $\gamma_0$, $\gamma_0'$ and $\gamma$ are 
temperature--dependent~\cite{BDV} friction coefficients.
Taking for example $R=0.0896~\rm cm$, we find that
$\Delta z/R \approx 13.4$,
$7.8$, $4.7$, $2.1$ and $1.3$ respectively at $T=1.3$, $1.5$, $1.7$, $1.9$,
$2.1$ and $2.15~\rm K$, in contradiction with the experiments that 
$\Delta z/R>20$.

The aim of this paper is therefore to go beyond the $T=0$ calculation of 
Ref.~\cite{Wacks-PoF}, and explore the effect of the friction.

\section{Method}

Following the method of Schwarz \cite{Schwarz}, we model vortex lines 
as space curves $\bs(\xi,t)$ of circulation $\kappa$ 
(where $t$ is time and $\xi$ is arc length)
which move according to

\begin{equation}
\label{eq:Schwarz}
\frac{ds}{dt} = \bv_s + \alpha \bs' \times (\bv_n - \bv_s) 
- \alpha' \bs' \times [\bs' \times (\bv_n - \bv_s)].
\end{equation}

\noindent
where $\alpha$ and $\alpha'$ are
temperature--dependent friction coefficients \cite{BDV,DB}, $\bv_n$ is
the normal fluid velocity, $\bs'=d\bs/d\xi$ is the unit tangent 
vector to the line at $\bs$, and 

\begin{equation}
\bv_s(\bs)=
-\frac{\kappa}{4\pi} \oint_{\mathcal{L}} 
\frac{(\bs - \br)}{|\bs - \br|^3} \times \bdr,
\label{eq:BS}
\end{equation}

\noindent
where the line integral extends over the entire vortex 
configuration $\mathcal{L}$. The technique to discretize the vortex lines,
desingularize the Biot--Savart integral~(\ref{eq:BS}) and perform
vortex reconnections when two vortex lines come very close to each
other has been already described in the literature
\cite{Baggaley-cascade,Baggaley-fluctuations,Baggaley-reconnections}.

The initial condition at $t=0$ is a bundle of $N$ coaxial concentric
vortex rings set within a torus of small radius $a$ and large radius $R$.
For large $N$, the rings are arranged so that they form a triangular lattice
on the cross--section of the torus \cite{Wacks-PoF}. This choice constraint
the choice of $N$ to ``hexagonal'' numbers $N=1,3,7,19, \cdots$.

In all numerical simulations, the distance between the discretization 
points along vortex lines is algorithmically kept between the values
$\Delta \xi$ and $\Delta\xi/2$, where $\Delta \xi$ is the parameter which
determines the numerical resolution. Typically we set 
$\Delta \xi$ so that $\ell/\Delta \xi \approx 10$ where $\ell$ is the
intervortex distance; for example, results for 
the $N=3$ bundle described below are obtained for
$R=0.06~\rm cm$, $\ell=0.015~\rm cm$
and $\Delta \xi=0.00149~\rm cm$, but we have done checks with smaller values
of $\Delta \xi$, e.g. $\Delta \xi=0.00075~\rm cm$, and found that the 
velocity of the bundle does not change. 

The computational cost of the 
Biot--Savart integrals (\ref{eq:BS}) scales with the square of the
number of discretization points; an increase of discretization points
also requires a smaller time step. Thus practical computing limitations
prevent us from simulating bundles with large $N$.

\section{Results}

First we recover the previous results \cite{Wacks-PoF}
corresponding to $T=0$ (no friction). The governing equation is
Eq.~(\ref{eq:Schwarz}) with $\alpha=\alpha'=0$.
For example, Fig.~(\ref{fig:1})
shows the stable, periodic leapfrogging motion of $N=3$ vortex 
rings around each other. In turn, the vortex ring at the back of the bundle
shrinks in size and goes inside 
the other two rings, speeding up and going ahead of them; then
it grows in size, slows down, and goes above and around the other rings,
falling behind them. This motion (a generalization of the well--known
leapfrogging of two vortex rings) repeats in
periodic fashion. It is interesting to notice that the vortex rings
remain circular -- no Kelvin waves develop.

Secondly, we study the effect of the friction on the motion of
the bundles.  The governing equation is
Eq.~(\ref{eq:Schwarz}) with nonzero $\alpha$ and $\alpha'$ but with
$\bv_n=0$ (normal fluid at rest).  We find that, in general,
a vortex bundle at finite temperatures in such stationary normal fluid
tends to lose its constituent vortex rings one by one, until it vanishes.
To illustrate this effect, we choose the same $N=3$ vortex
configuration of Fig.~(\ref{fig:1}), and show what happens to it at
at $T=2.02~\rm K$ in Fig.~(\ref{fig:2}). The ring which is inside the
other two rings
tends to shrink too much, speeding up and going ahead of them,
then escaping the bundle and vanishing away.

Larger bundles ($N>3$) suffer the same fate, which is 
not consistent with Borner's observation
that bundles travel in a stable way for a distance much larger
than their size. The question is: what stabilizes the bundles
in the experiments ?

We argue that the answer is the viscous normal fluid, which is not at rest,
but is pushed out of the hole together with the superfluid,
and must roll up in the classical form of an ordinary vortex ring. 
In fact Borner \etal measured the circulation 
of this normal fluid ring and found that it is the same
circulation of the superfluid vortex bundle.
To model the normal fluid ring, we define the following toroidal, Rankine
vortex flow $\bv_n$. 
At each instant, let the origin of our reference
frame be the (moving) centre of the superfluid vorticity and $z$ be
the direction of propagation of the bundle. In each cross section of the 
torus, let $r$ be the radial distance away from the axis around the torus.
The normal fluid velocity $\bv_n$ has two components. 
The first component is a uniform flow along $z$ with the same speed
of the centre of the superfluid vorticity. The second component, tangential
around the torus, is equal to a uniform solid body rotation
(proportional to $r$) inside the torus, and an irrotational flow 
(inversally proportional to $r$) with 
circulation $N \kappa$ outside the torus \cite{note1}. 
To accommodate distortions of the core of
superfluid bundle (which is initially circular, but 
tends to become D--shaped during the evolution)
and the dissipative nature of the normal fluid vortex
ring (which we expect to spreads spatially), the transition from solid body 
rotation and irrotational flow is not at $r=a$, but at $r=2a$.

The model is simple but gives results in agreement with the
experiments. Fig.~(\ref{fig:3}) shows the same
$N=3$ superfluid vortex bundle at $T=2.02~\rm K$ with $\bv_n$ defined
as above. It is apparent that the bundle moves in a stable way.

The result holds true for bundles with $N>3$. Table~(\ref{tab:1}) 
summarizes the initial conditions of our numerical simulations, listing
the large radius $R$ and the small radius $a$ of the torus,
their ratio $R/a$, the initial vortex length $\lambda_0$
(which is not exactly equal to $2 \pi R N$ because the rings within the 
torus have different radii), the initial average curvature (which, for
the same reason, is not exactly equal to $1/R$), and the initial 
intervortex distance $\ell_0$.
These initial conditions are chosen to best fit Borner's experiment 
with the same $\ell_0$, see the discussion in Ref.~ \cite{Wacks-PoF}.

Tables~({\ref{tab:2}) and ~(\ref{tab:3}) show the results of our
numerical simulations respectively at $T=0$ (no friction) 
and $T=2.02~\rm K$ (with friction and normal fluid vortex ring $\bv_n$). 
The simulations are stopped when a vortex bundle has travelled a 
distance of the order of 10 diameters, as observed in the
experiments.
The tables list the time $t$ at which we stop the simulation, 
the distance travelled $\Delta z$ by the centre of vorticity
of each bundle in terms of the initial diameter $2 R$, 
the vortex length $\Lambda$ at time $t$ in terms of
the initial vortex length $\Lambda_0$, the average curvature 
${\bar c}$ at time $t$ in terms of the initial
average curvature ${\bar c}_0$, and the speed $v$ of the vortex 
bundle at time $t$ in terms of the initial speed $v_0$.

The main result is that, in the presence of the normal fluid vortex ring, 
a superfluid bundle travels a significant
distance despite the friction, in agreement with the experiment 
of Borner \etal.  

We find that, with or without friction, the smaller bundles (small $N$)
tend to remain
circular (at least within the time--scales of our simulations);
the larger bundles (large $N$) 
develop instabilities which induce vortex 
reconnections,  changing the actual number of vortex rings in the bundle,
and causing Kelvin waves. 
Interestingly, vortex reconnections have no significant effect on the 
coherence of a bundle -- a turbulent vortex bundle only seems to travel
somewhat slower. 

More precisely, the $N=7$ bundle has the first reconnection
at $t=26.98~\rm s$ for $T=0$ and at about the same time ($t=24.06~\rm s$)
for $T=2.02~\rm K$. The fact that the first reconnection
in the evolution of the $N=19$ bundle
at $T=0$ occurs much sooner ($t=19.89~\rm s$)
than at $T=2.02~\rm K$ ($t=45.76~\rm s$).

The main difference between the appearance of the vortex bundles
at $T=0$ and $T=2.02~\rm K$ is 
the wavelengths which seem excited: low temperatures favour short waves,
high temperatures long waves, as shown in Figs.~(\ref{fig:4}) and 
(\ref{fig:5}) respectively. The increase of average curvature reflects
this effect. This is consistent with the observation made by
Tsubota {\it et al.} \cite{Araki}
that, at low temperatures, turbulent vortex tangles tend to be more wiggly.

\section{Conclusion}

Practical computer limitations
prevents us from studying vortex bundles with thousands of rings as
in the experiments. Nevertheless, exploring in detail what happens to
small bundles gives us insight into the physics of the
problem. 

Our results suggest that
bundles of superfluid vortex rings
can travel coherently a significant distance, at least one order
of magnitude larger than their diameter, in agreement with 
experimental observations. The effect seems temperature independent.
The normal fluid vortical structure which is
generated by the piston--cylinder set up has been observed to
to move along the superfluid vortical structure. The coexistence
of superfluid and normal fluid structures
effectively inhibits the friction between the 
superfluid vortices and the normal fluid from dissipating the
bundles, and this effect explains the experimental observation
of Borner {\it et al.}
\cite{Borner1981,Borner1983,Borner1985}
that large--scale vortex rings remain stable at nonzero temperatures.

To put the large--scale vortex ring experiments in a wider context,
it is worth recalling a related problem: the
``spontaneous'' appearance of bundles of vortices in superfluid turbulence
(opposed to the ``forced'' generation of bundles of vortices by the 
piston--cylinder arrangement described here). Missing any direcy experimental
observation, the existence and the non--existence of such bundles 
\cite{Volovik2003,Morris2008,Kivotides2011,SkrbekSreeni2012,Nemirovskii2013}
or of partial polarization of vortex lines
\cite{RocheBarenghi2008,BaggaleyLaurieBarenghi2012}, 
has been discussed in the literature, particularly with respect to the
Kolmogorov spectrum \cite{Nemirovskii2014,BarenghiLvovRoche2014}. 
In this context, the vortex rings generated by the 
piston--cylinder setup provide a ``forced'' but controlled method to study the 
coupling of normal fluid and superfluid.

Finally, we notice that the detailed mechanism of generation of
the double (normal fluid and superfluid) vortex ring structure at the 
hole of the cylinder is an interesting problem of two--fluid hydrodynamics
which would be worth studying.

\newpage

\begin{figure}[ht!]
\includegraphics[width=0.30\textwidth]{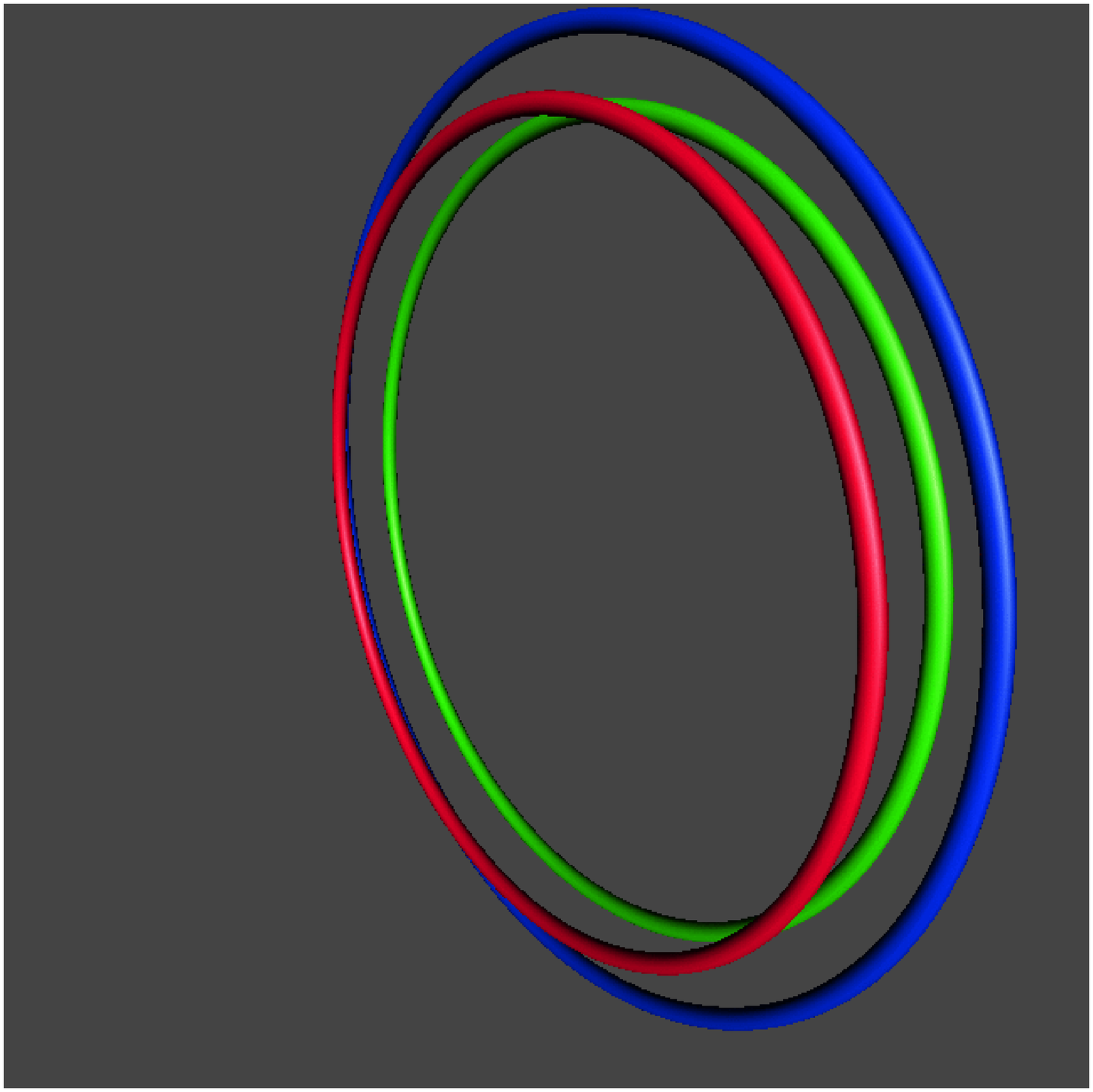}
\includegraphics[width=0.30\textwidth]{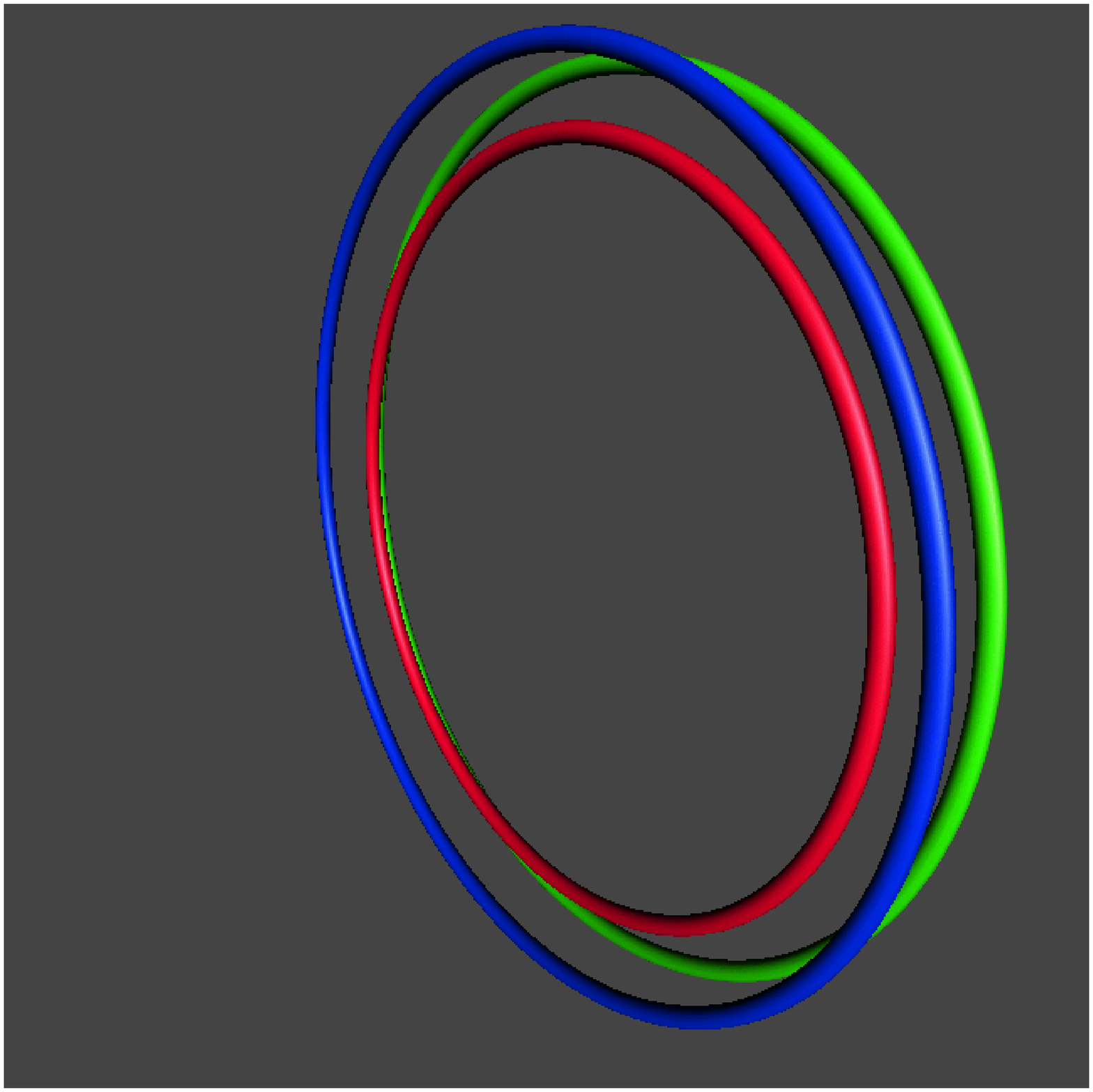}
\includegraphics[width=0.30\textwidth]{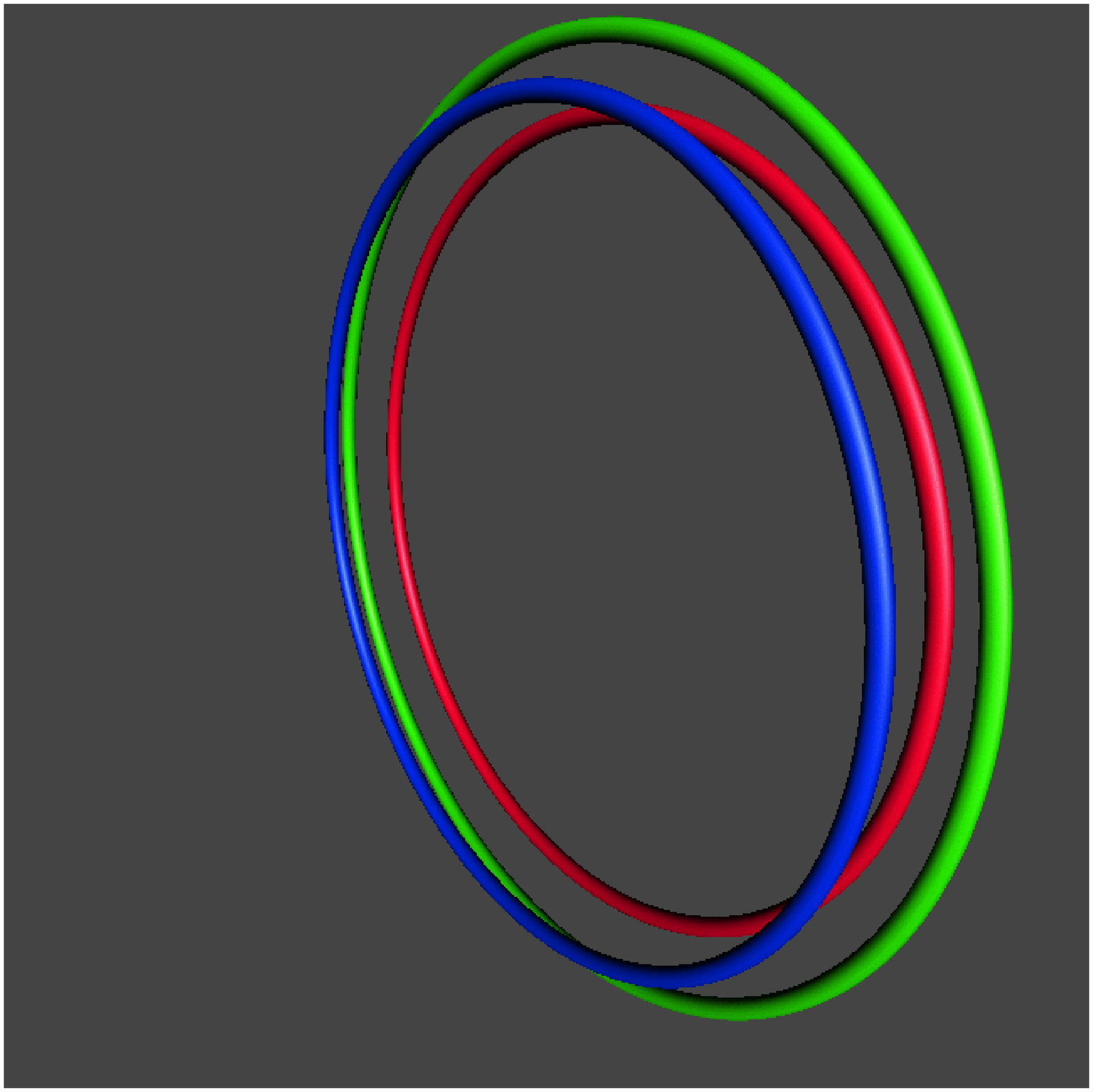}
\caption{\label{fig:1}(Color online). Motion of $3$ vortex rings 
(initial condition: $R=0.0896~{\rm cm}$, $a=0.0075~{\rm cm}$)
at zero temperature (no friction).
Left: $t=0$; middle: $t=3.6$; right: $t=7.2~{\rm s}$. 
The vortex bundle is stable, and each ring leapfrogs around the others.
Each vortex line is represented by a tube\cite{knotplot} 
whose thickness if
for the sake of visualization only (vortex lines have infinitesimal
thickness).
}
\end{figure}
\begin{figure}[ht!]
\includegraphics[width=0.30\textwidth]{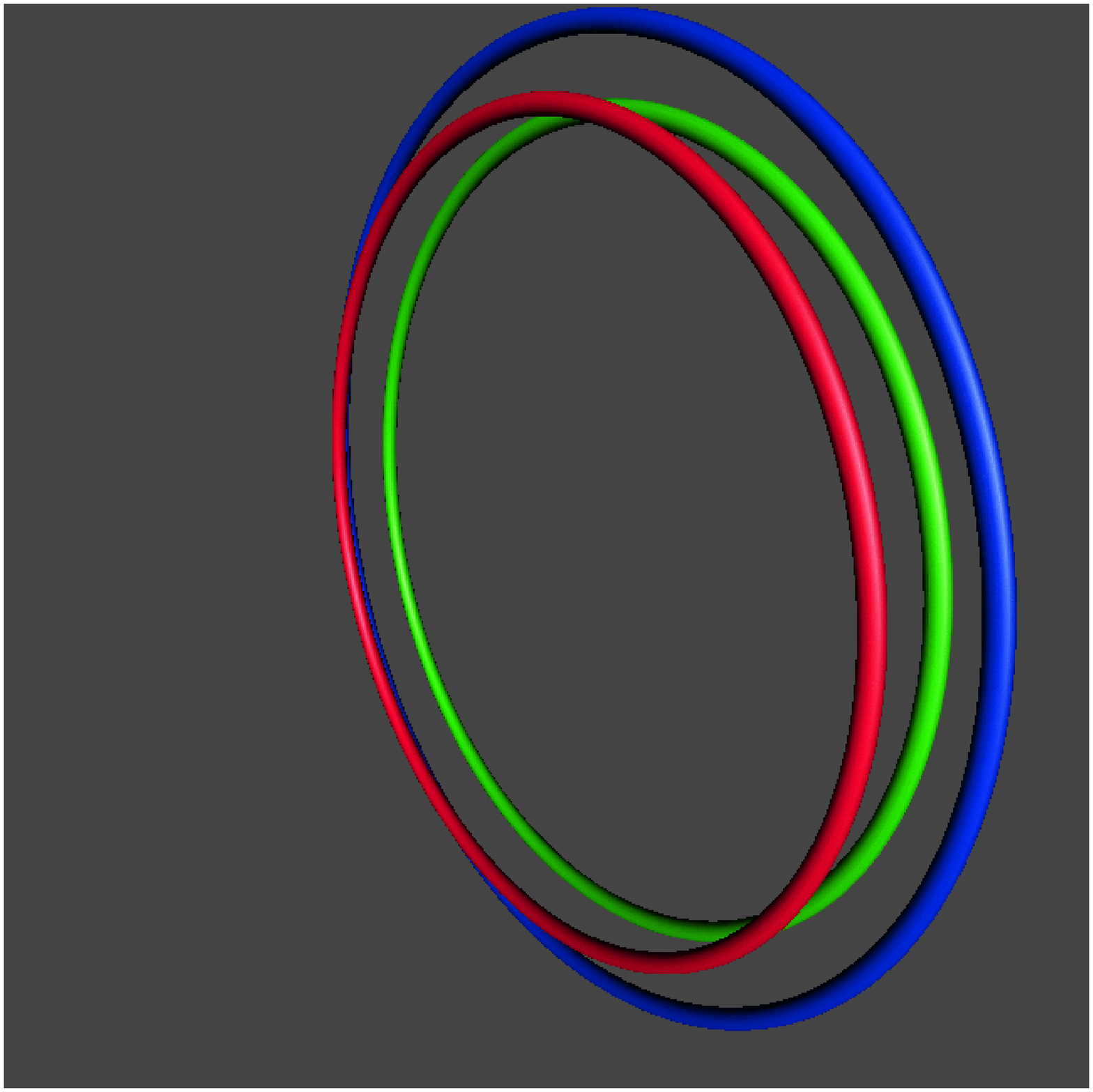}
\includegraphics[width=0.30\textwidth]{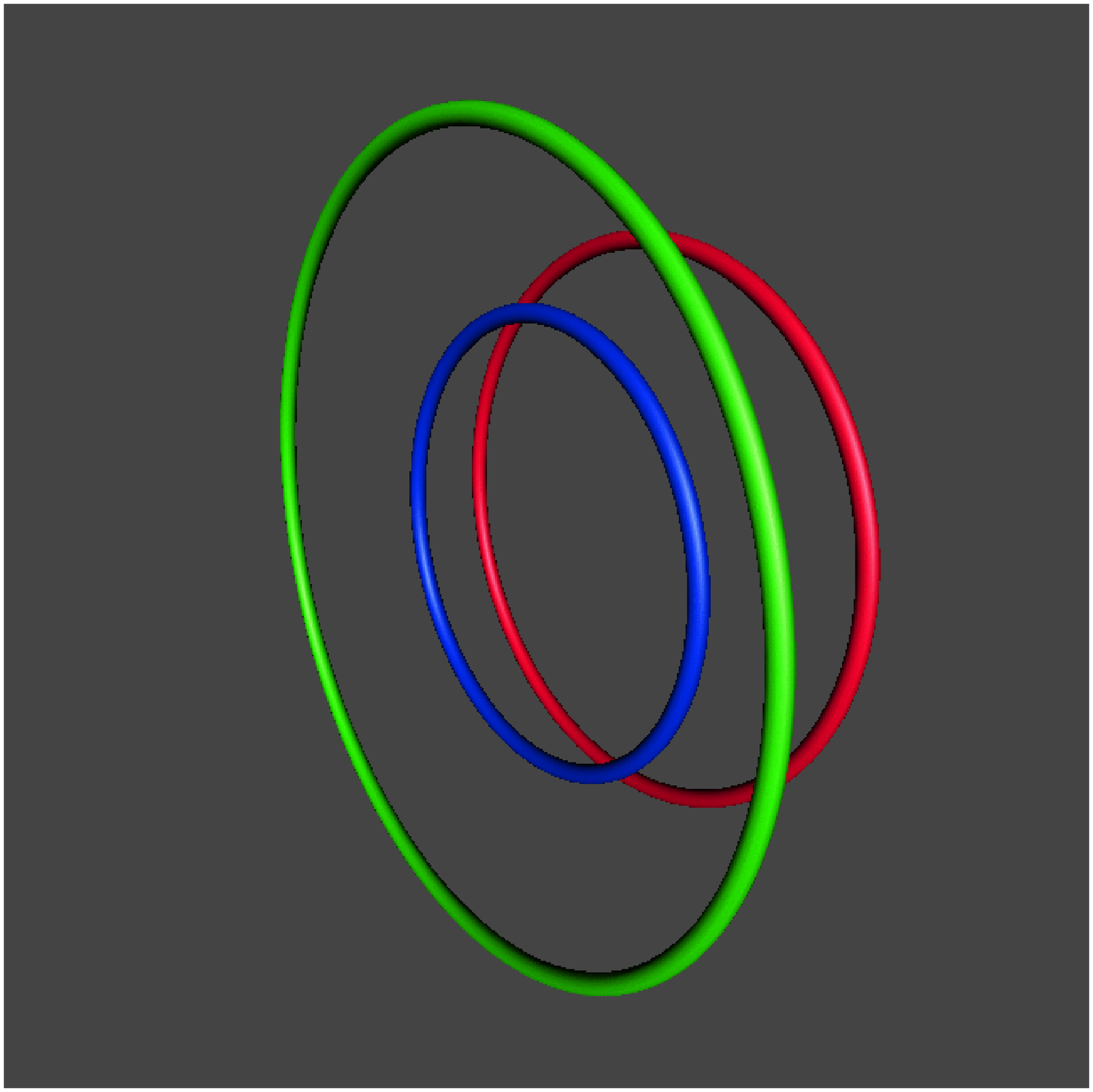}
\includegraphics[width=0.30\textwidth]{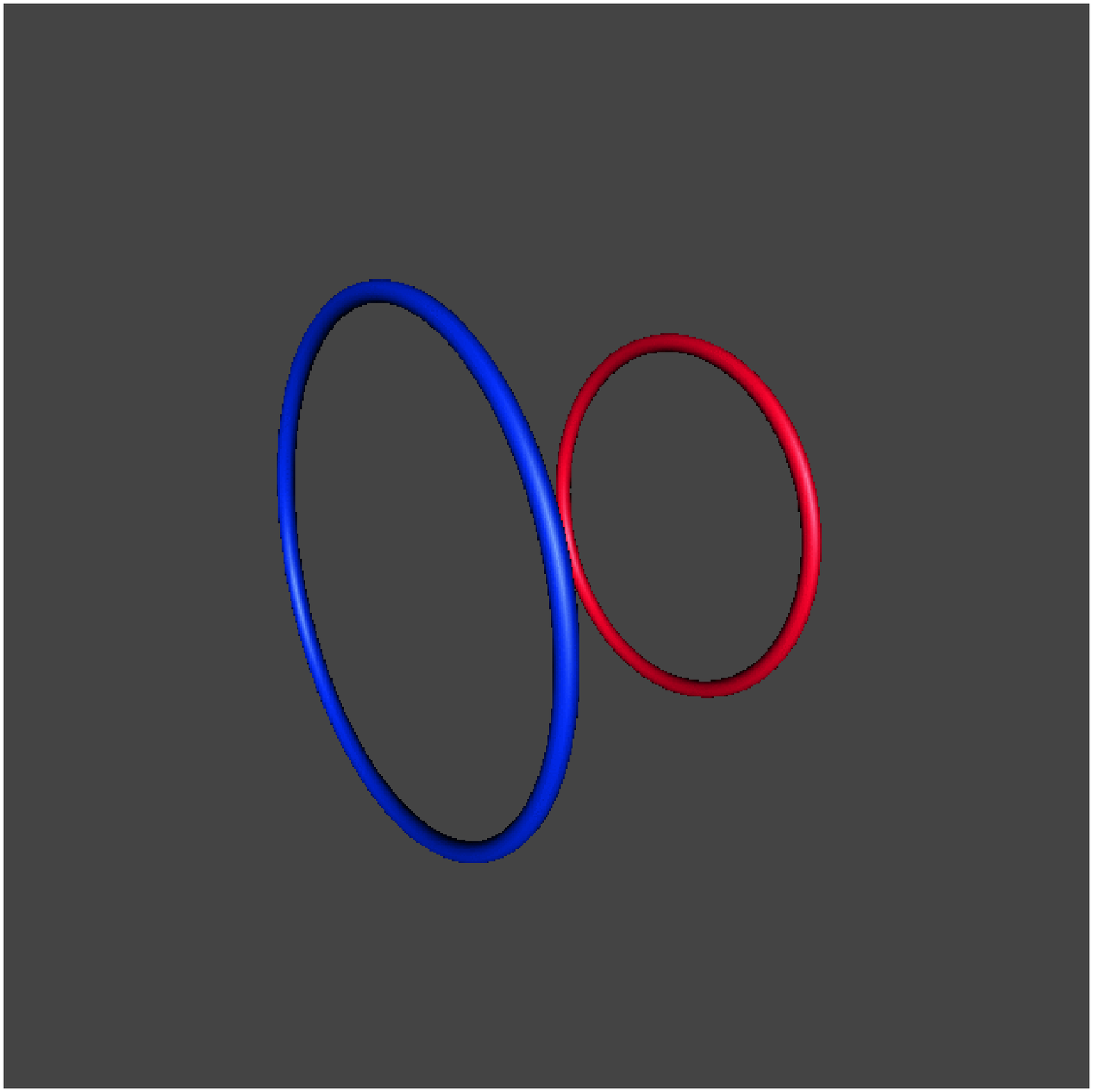}
\caption{\label{fig:2} (Color online). Motion of $3$ vortex rings 
in the presence of friction at $T=2.02~\rm K$
with $\bv_n=0$.
The initial condition is the same as in Fig.~(\ref{fig:1}).
Left: $t=0$; middle: $t=3.6$; right: $t=7.2~{\rm s}$. 
The vortex bundle decays: one by one, all vortex rings shrink and vanish
on the axis of propagation.
}
\end{figure}
\begin{figure}[ht!]
\includegraphics[width=0.30\textwidth]{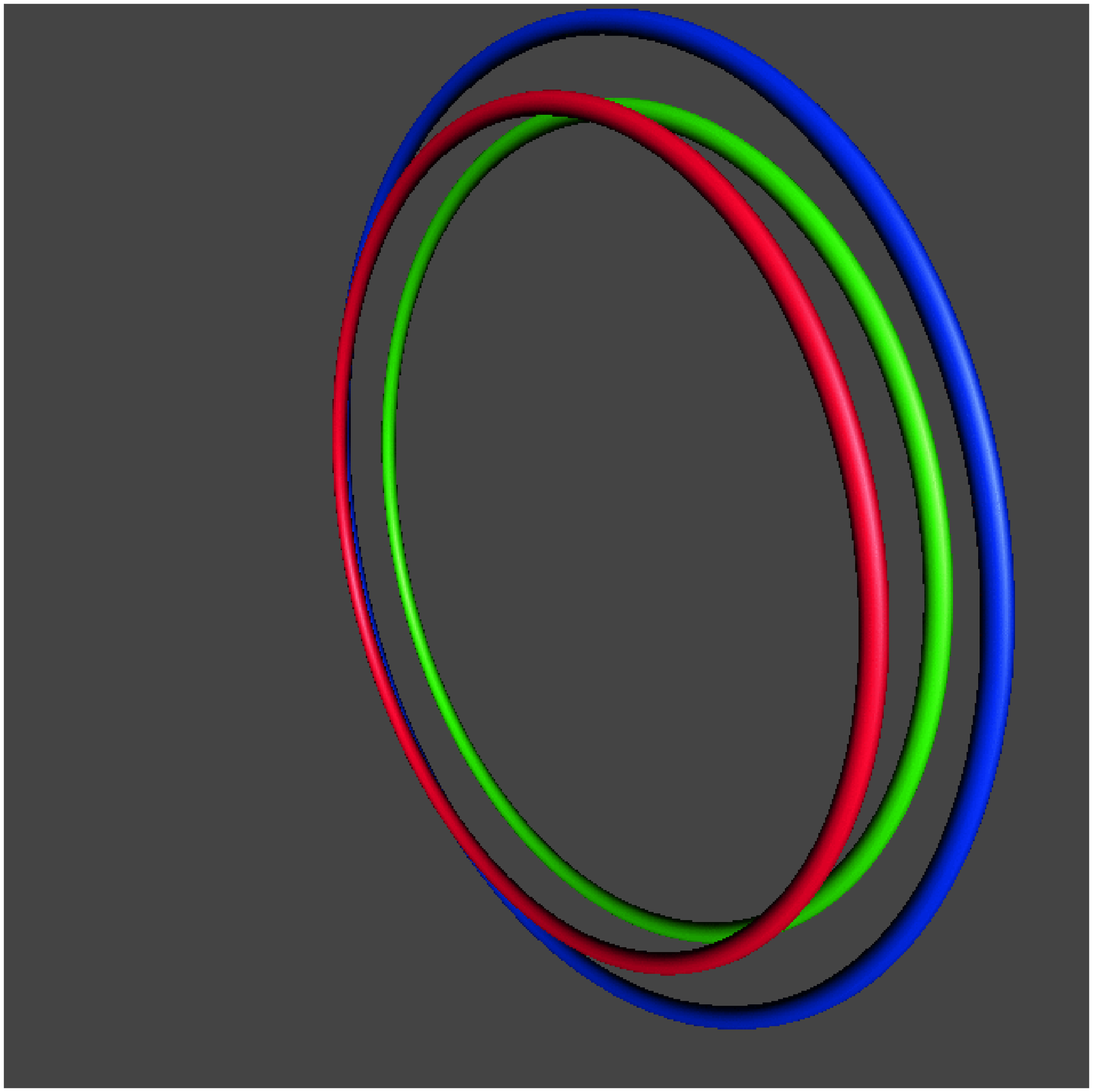}
\includegraphics[width=0.30\textwidth]{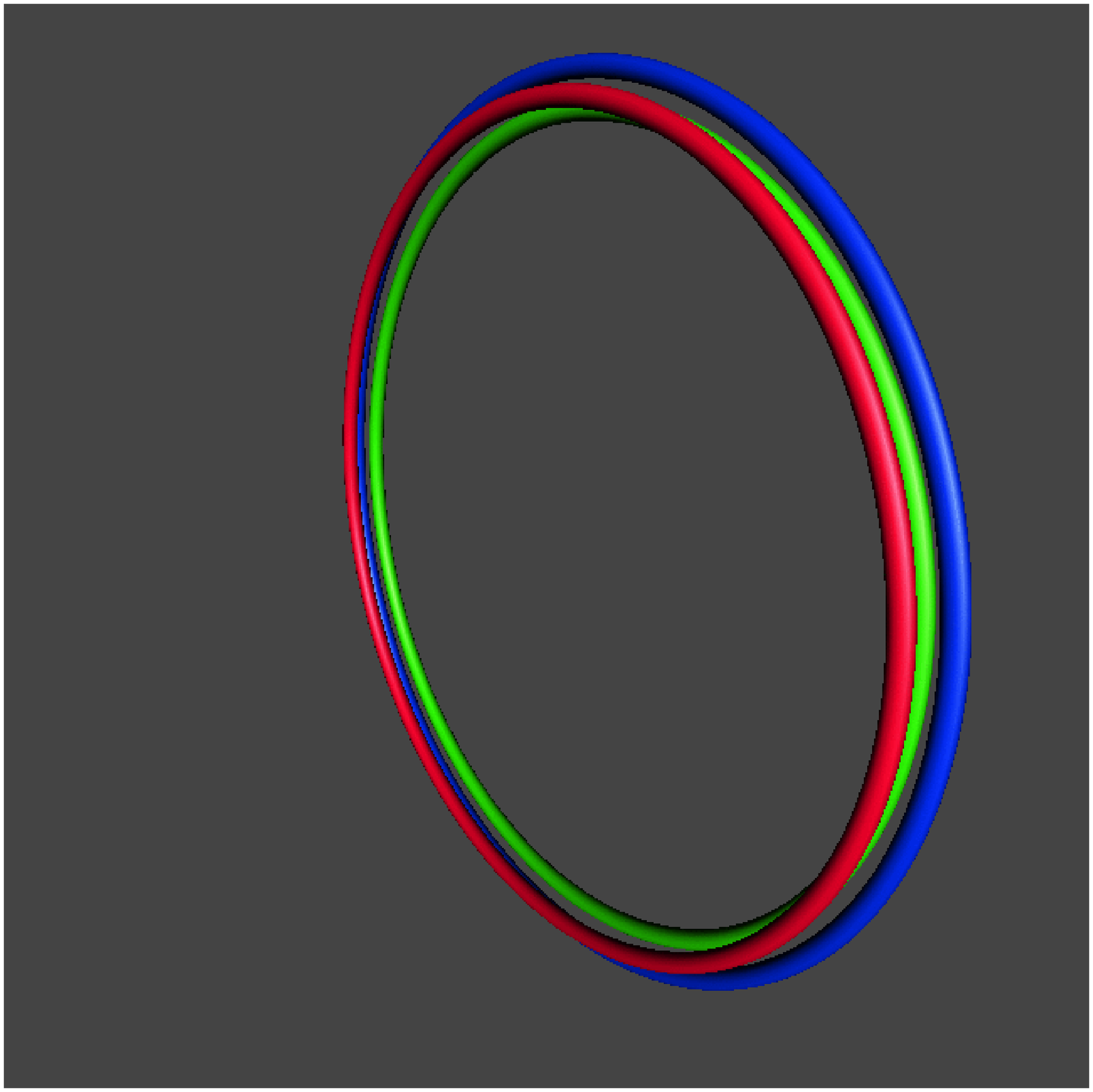}
\includegraphics[width=0.30\textwidth]{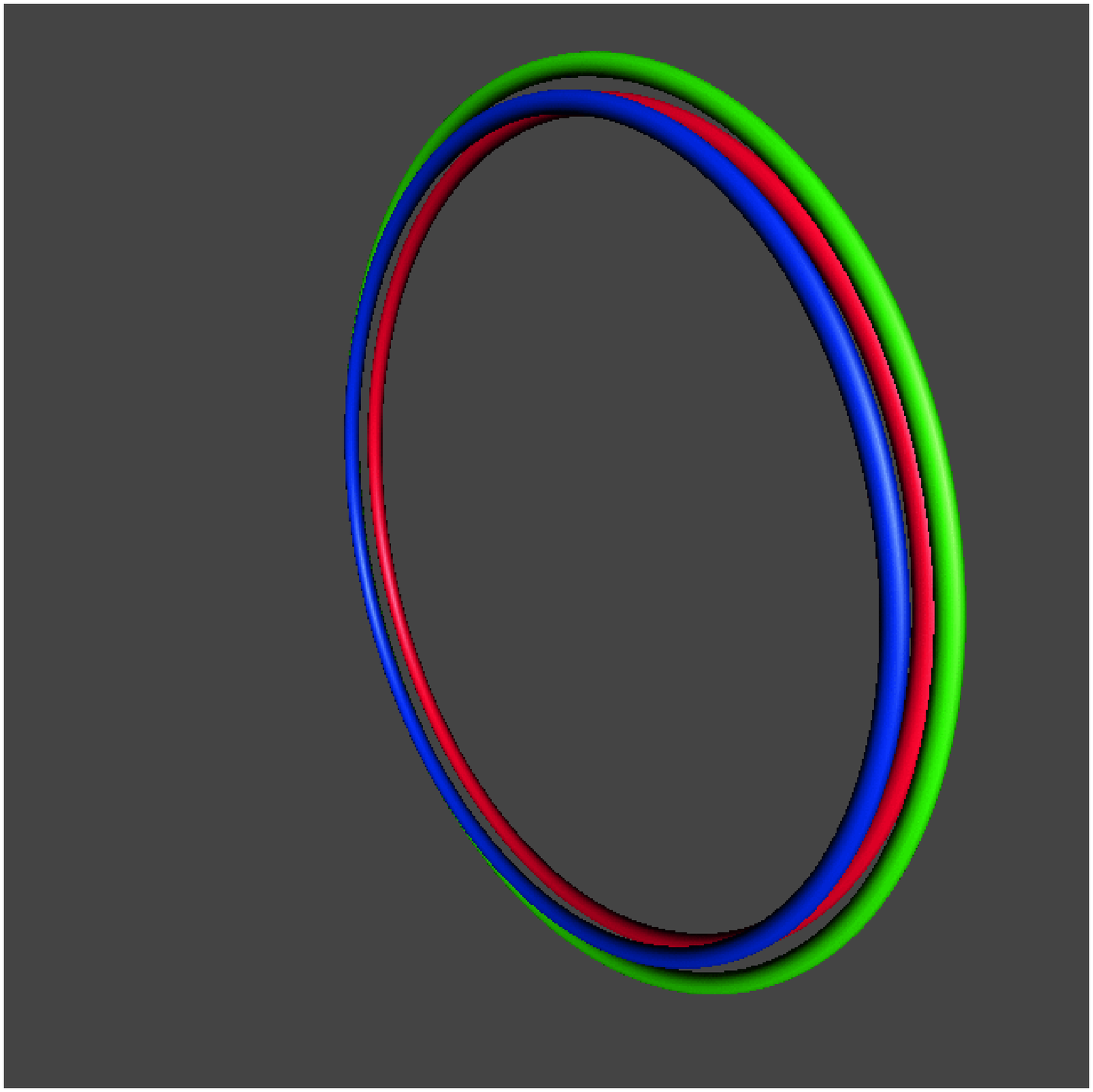}
\caption{\label{fig:3} (Color online). Motion of $3$ vortex rings in the
presence of friction at $T=2.02~\rm K$ and normal fluid ring
($\bv_n\neq 0$).
The initial condition is the same as in Fig.~(\ref{fig:1}). 
Left: $t=0$; middle: $t=3.6$; right: $t=7.2~{\rm s}$. 
It is apparent that the 
superfluid vortex bundle moves in a stable way as in the
absence of friction (Fig.~(\ref{fig:1});  the individual rings leapfrog
around each other.
}
\end{figure}
\begin{figure}[ht!]
\includegraphics[width=0.30\textwidth]{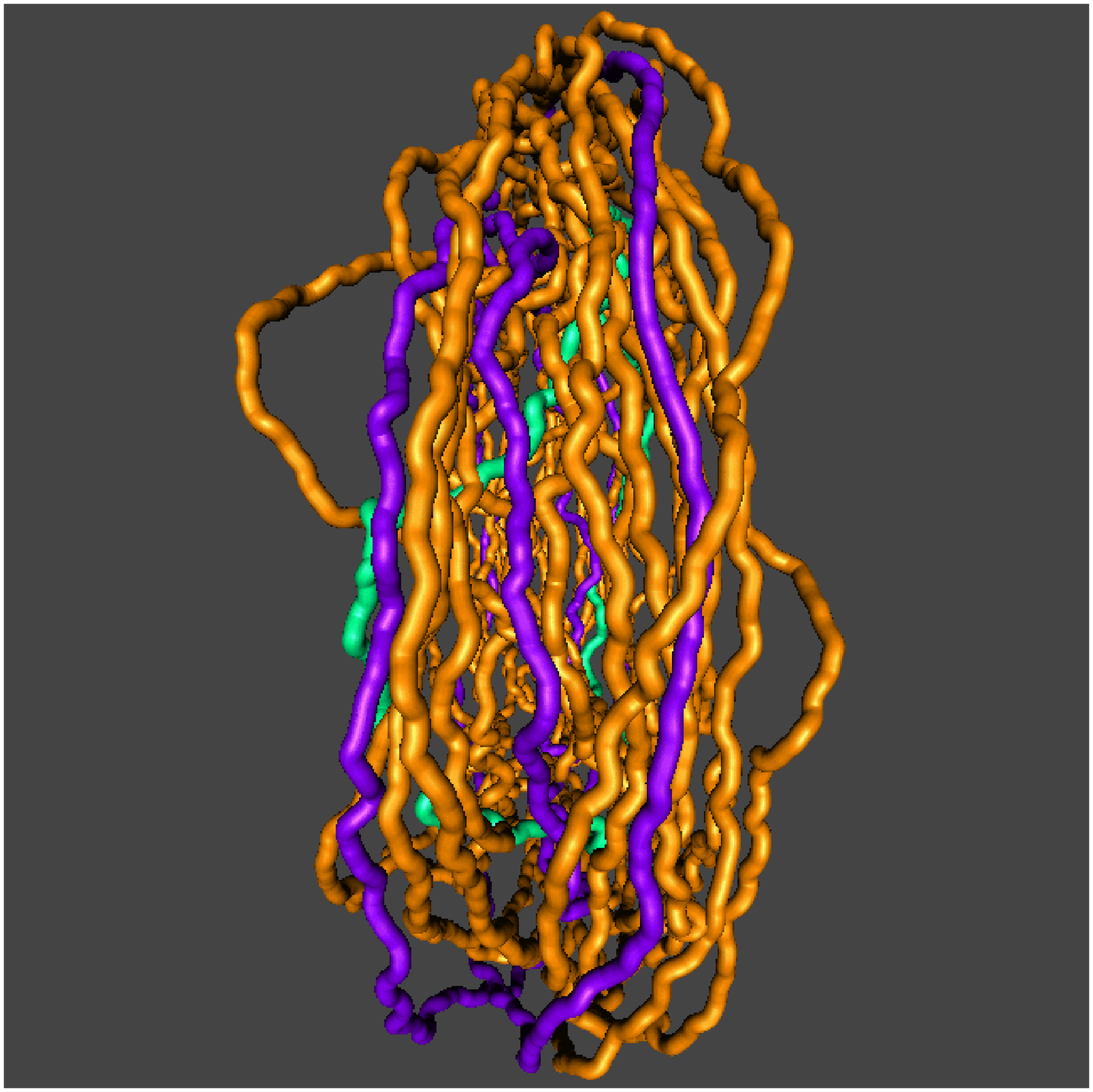}
\includegraphics[width=0.30\textwidth]{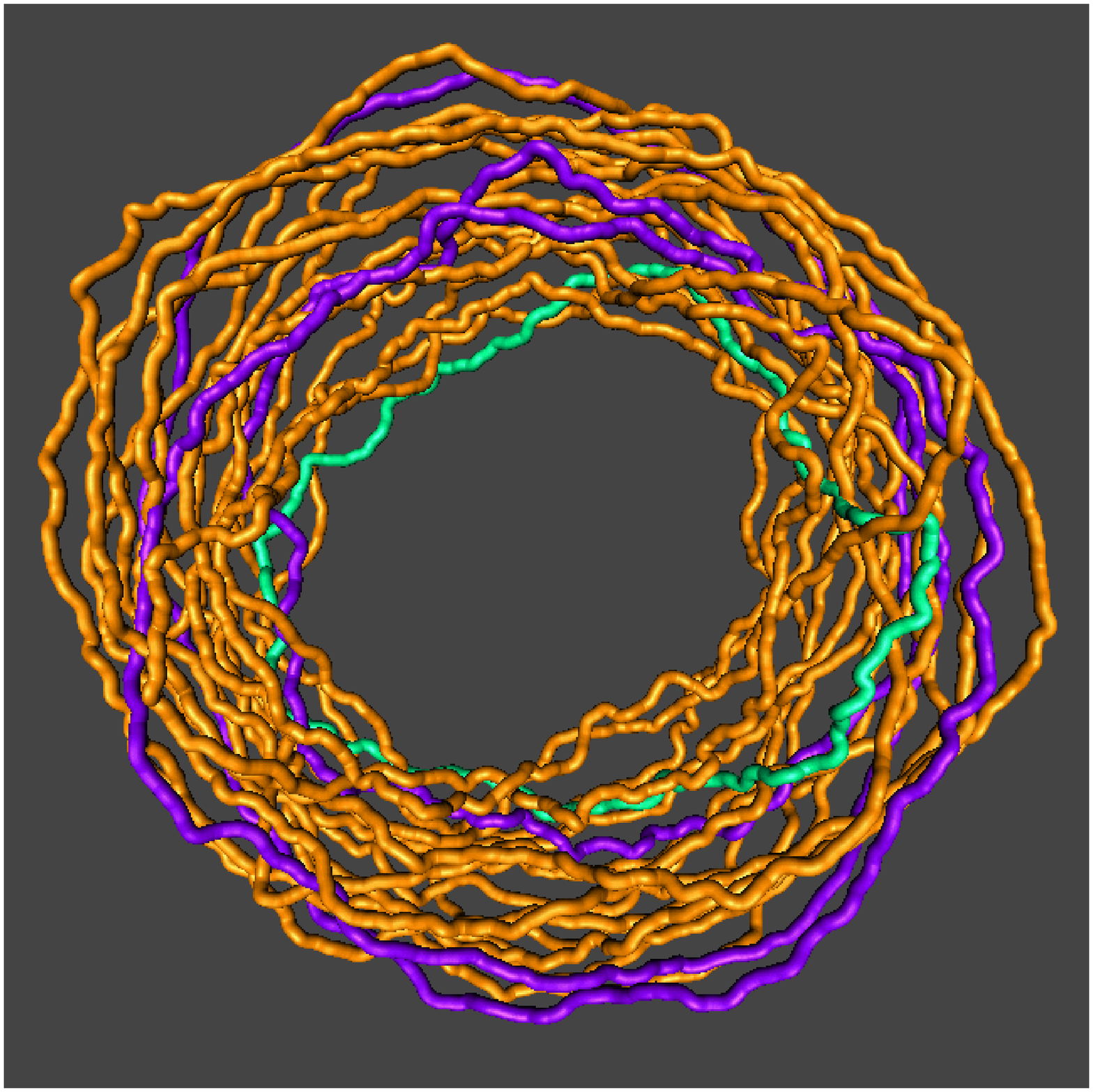}
\caption{\label{fig:4} 
(Color online).  Vortex bundle at $T=0$ (no friction).
Top: Side (left) and rear (right) view
of vortex bundle with $N=19$ rings at time $t=40~\rm s$.
}
\end{figure}
\begin{figure}[ht!]
\includegraphics[width=0.30\textwidth]{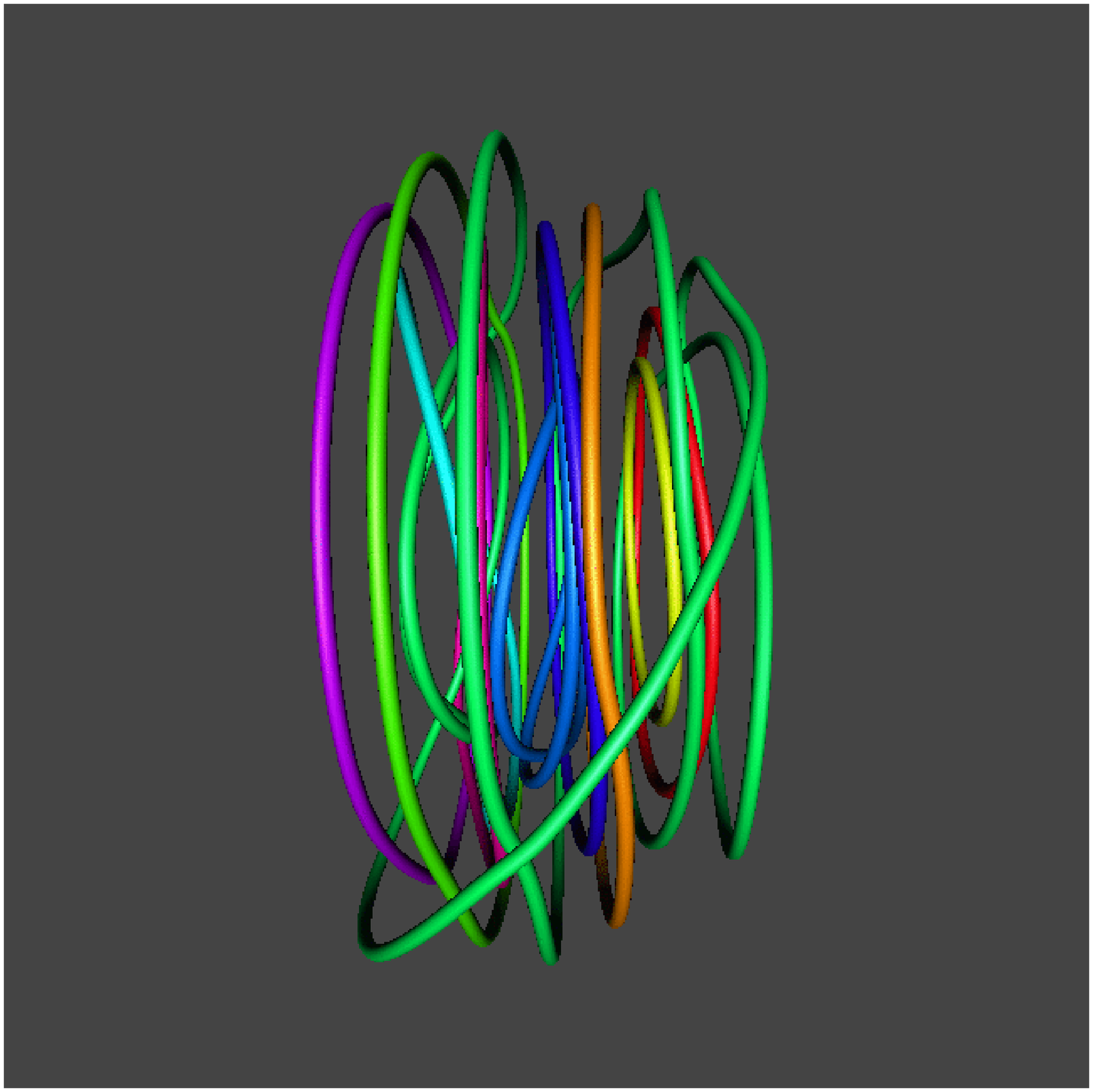}
\includegraphics[width=0.30\textwidth]{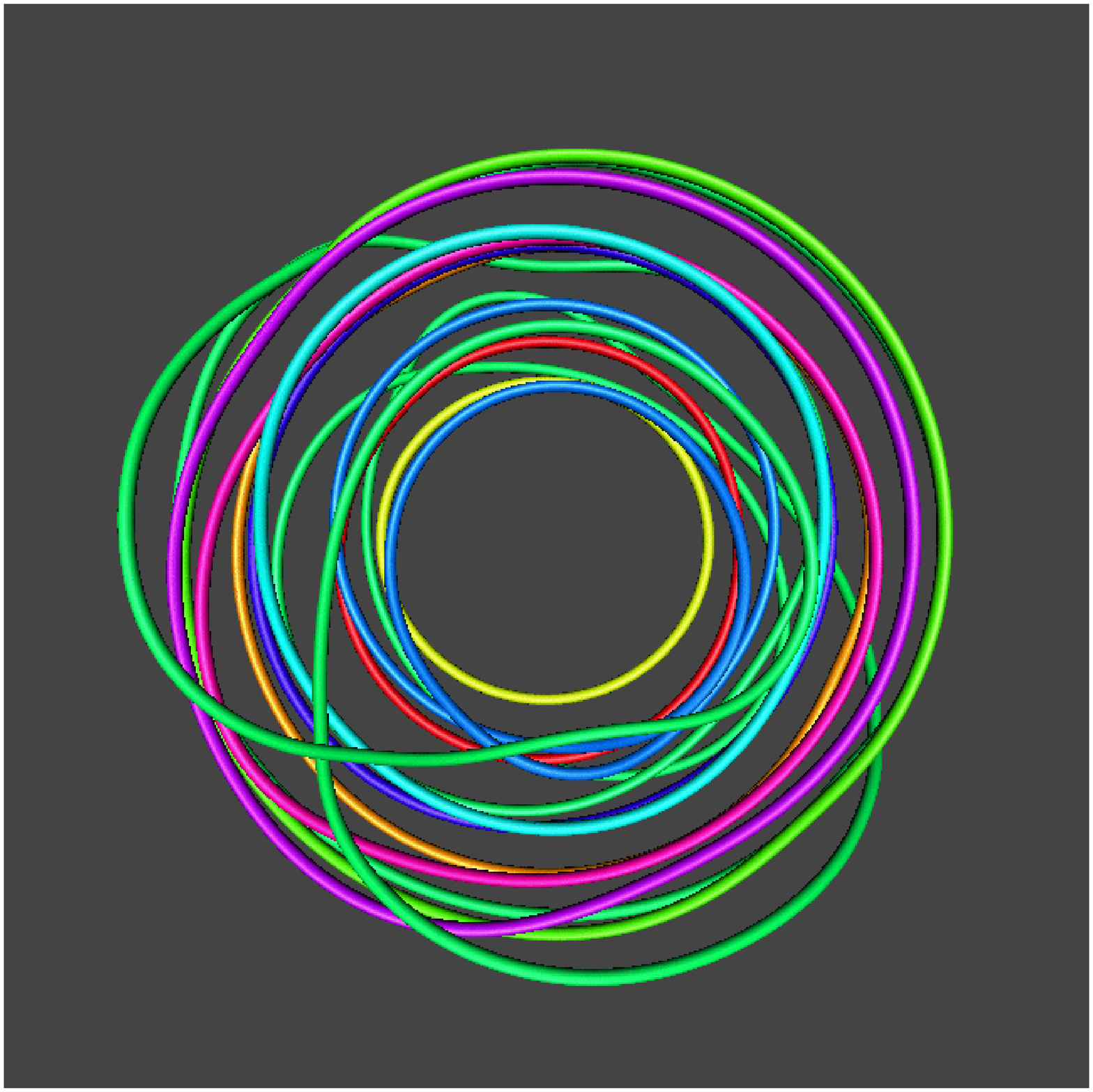}
\caption{\label{fig:5} 
(Color online).
Vortex bundle at $T=2.02~\rm K$ in the presence of friction and 
normal fluid ring $\bv_n$.
Top: Side (left) and rear (right) view
of vortex bundle with $N=19$ rings at time $t=80~\rm s$.
}
\end{figure}
\newpage
\begin{table}
\centering
\begin{tabular}{|c||c|c|c|c|c|c|c|}
\hline
$N$ & $R$         & $a$         & $R/a$   & $\Lambda_0$  & ${\bar c_0}$      & $\ell_0$   & $v_0$\\
    & ${\rm cm}$  & ${\rm cm}$  &         & ${\rm cm}$   & ${\rm cm^{-1}}$   & ${\rm cm}$ & ${\rm cm/s}$\\ 
\hline
 2  & 0.06        & 0.075       & 8.0      & 0.754        & 17.19             & 0.015     & 0.0279\\
 3  & 0.06        & 0.00866     & 6.92     & 1.131        & 17.11             & 0.015     & 0.0310 \\
 7  & 0.06        & 0.015       & 4.0      & 2.639        & 18.08             & 0.015     & 0.0450 \\
 19 & 0.12        & 0.013       & 4.0      &14.326        & 10.87             & 0.015     & 0.0471\\
\hline
\end{tabular}
\caption{Initial conditions.}
\label{tab:1}
\end{table}
\begin{table}
\centering
\begin{tabular}{|c||c|c|c|c|c|c|}
\hline
N & $t$          & $\Delta z/(2R)$   & $\Lambda/\Lambda_0$   & ${\bar c}/{\bar c_0}$    & $v/v_0$ \\ 
  & ${\rm s}$    &                &                       &                          &        \\
\hline
2 & 50            &  11.11       & 1.0                    & 0.97                     & 0.92   \\ 
3 & 40            &  10.24       & 1.0                    & 1.0                      & 1.0    \\
7 & 30            &  10.91       & 1.1                    & 3.8                      & 0.84   \\
19& 60            &  10.38       & 1.6                    & 18.4                     & 0.59   \\
\hline
\end{tabular}
\caption{Evolution at $T=0$ (no friction).}
\label{tab:2}
\end{table}

\begin{table}
\centering
\begin{tabular}{|c||c|c|c|c|c|c|}
\hline
N & $t$          & $\Delta z/D$   & $\Lambda/\Lambda_0$   & ${\bar c}/{\bar c_0}$    & $v/v_0$ \\ 
  & ${\rm s}$    &                &                       &                          &        \\
\hline
2 & 50            & 11.87         & 0.97                       &  1.02               & 0.99      \\ 
3 & 40            & 11.37         & 0.96                       &  1.07               & 1.10       \\
7 & 30            & 10.13         & 1.52                       &  1.18               & 0.69       \\
19& 60            & 10.21         & 0.80                      &  1.08               & 0.64       \\
\hline
\end{tabular}
\caption{Evolution at $T=2.02~\rm K$ in the presence of friction and $\bv_n$.}
\label{tab:3}
\end{table}


\begin{thebibliography}{99}
\bibitem{note1}
This assumption is justified by our finding \cite{Wacks-PoF}
that the translational speed of the superfluid bundle is approximately
equal to the speed of a single vortex ring with circulation $N \kappa$.
\end{thebibliography}
\end{document}